\documentclass[12pt]{article}
\usepackage{epsfig,graphicx}
\usepackage{fpcp03}

\newcommand{\lsim}{
\mathrel{\hbox{\rlap{\hbox{\lower4pt\hbox{$\sim$}}}\hbox{$<$}}}}

\newcommand{\gsim}{
\mathrel{\hbox{\rlap{\hbox{\lower4pt\hbox{$\sim$}}}\hbox{$>$}}}}

\begin{document}

%%%%%%%%%%% CERN Titlepage %%%%%%%%%%%

\begin{titlepage}

\begin{flushright}
CERN-TH/2003-149\\
hep-ph/0307033
\end{flushright}

\vspace{1cm}
\begin{center}
\Large\bf CP Violation and New Physics in \boldmath $B_s$\unboldmath\ Decays
\end{center}

\vspace{1.0cm}
\begin{center}
{\large Robert Fleischer}\\[0.1cm]
{\sl Theory Division, CERN, CH-1211 Geneva 23, Switzerland}
\end{center}

\vspace{1.2cm}

\begin{center}
{\large {\bf Abstract}}
\end{center}

\vspace{0.3cm}

\begin{quotation}
\noindent
The $B_s$-meson system is a key element in the $B$-physics programme of hadron 
colliders, offering various avenues to explore CP violation and to search for 
new physics. One of the most prominent decays is $B_s\to J/\psi\phi$, the 
counterpart of $B_d\to J/\psi K_{\rm S}$, providing a powerful tool to
search for new-physics contributions to $B^0_s$--$\overline{B^0_s}$ mixing. 
Another benchmark mode is $B_s\to K^+K^-$, which complements 
$B_d\to\pi^+\pi^-$, thereby allowing an extraction of the angle $\gamma$ of 
the unitarity triangle that is sensitive to new-physics effects in the 
QCD penguin sector. Finally, we discuss new methods to constrain and 
determine $\gamma$ with the help of $B_s\to D_s^{(\ast)\pm} K^\mp$ decays, 
which complement $B_d\to D^{(\ast)\pm}\pi^\mp$ modes. Since these 
strategies involve ``tree'' decays, the values of $\gamma$ thus extracted 
exhibit a small sensitivity on new physics.
\end{quotation}

\vspace{1.0cm}

\begin{center} 
{\sl Invited talk at FPCP 2003 -- Flavour Physics \& CP Violation\\
Ecole Polytechnique, Paris, France, 3--6 June 2003\\
To appear in the Proceedings}
\end{center}

\vfill
\noindent
CERN-TH/2003-149\\
July 2003

\end{titlepage}

\thispagestyle{empty}
\vbox{}
\newpage
 
\setcounter{page}{1}

%%% end CERN title page %%%%%%%%%%%%%

\Title{CP Violation and New Physics in $B_s$ Decays}
\bigskip

%+ \addcontentsline{toc}{chapter}{{\it Robert~Fleischer}}
%+ \index{author}{Fleischer, R.} 

%%%%%%%%%%%%%%%%%%%%%%%%%%%%%%%%%%%%%
% Label to flag the first page of your contribution
% Replace Perret by your name starting with a capital letter
%
\label{FleischerStart}

%%%%%%%%%%%%%%%%%%%%%%%%%%%%%%%%%%%%%
% Your name
%
\author{Robert Fleischer\index{Fleischer, R.} }

%%%%%%%%%%%%%%%%%%%%%%%%%%%%%%%%%%%%%
% Your address
%
\address{Theory Division, CERN\\
CH-1211 Geneva 23, Switzerland\\
}

\makeauthor\abstracts{
The $B_s$-meson system is a key element in the $B$-physics programme of hadron 
colliders, offering various avenues to explore CP violation and to search for 
new physics. One of the most prominent decays is $B_s\to J/\psi\phi$, the 
counterpart of $B_d\to J/\psi K_{\rm S}$, providing a powerful tool to
search for new-physics contributions to $B^0_s$--$\overline{B^0_s}$ mixing. 
Another benchmark mode is $B_s\to K^+K^-$, which complements 
$B_d\to\pi^+\pi^-$, thereby allowing an extraction of the angle $\gamma$ of 
the unitarity triangle that is sensitive to new-physics effects in the 
QCD penguin sector. Finally, we discuss new methods to constrain and 
determine $\gamma$ with the help of $B_s\to D_s^{(\ast)\pm} K^\mp$ decays, 
which complement $B_d\to D^{(\ast)\pm}\pi^\mp$ modes. Since these 
strategies involve ``tree'' decays, the values of $\gamma$ thus extracted 
exhibit a small sensitivity on new physics.
}

\section{Setting the Stage}\label{fleischer:sec-intro}
At the $e^+e^-$ $B$ factories operating at the $\Upsilon(4S)$ resonance, 
$B_s$ mesons are not accessible, i.e.\ their decays cannot be explored by 
the BaBar, Belle and CLEO collaborations. On the other hand, plenty of 
$B_s$ mesons will be produced at hadron colliders. Consequently, these 
particles are the ``El Dorado'' for $B$-decay studies at run II of the 
Tevatron \cite{fleischer-TEV-BOOK}, and later on at the 
LHC~\cite{fleischer-LHC-BOOK}. A detailed overview of the physics 
potential of $B_s$ mesons can be found in \cite{fleischer-RF-PHYS-REP}.

An important aspect of $B_s$ physics is the mass difference $\Delta M_s$,
which can be complemented with $\Delta M_d$ to determine the side 
$R_t\propto|V_{td}/V_{cb}|$ of the unitarity triangle (UT). To this end, we 
use that $|V_{cb}|=|V_{ts}|$ to a good accuracy in the Standard Model (SM), 
and require an $SU(3)$-breaking parameter, which can be determined, e.g.\ 
on the lattice. At the moment, only experimental lower bounds on 
$\Delta M_s$ are available, which can be converted into upper bounds on 
$R_t$, implying $\gamma\lsim 90^\circ$ \cite{fleischer-CKMyellow2002}. 
Once $\Delta M_s$ is measured, more stringent constraints on $\gamma$ 
will emerge. 

Another interesting quantity is the width difference $\Delta\Gamma_s$. 
While $\Delta\Gamma_d/\Gamma_d$ is negligibly small, where $\Gamma_d$ 
is the average decay width of the $B_d$ mass eigenstates, 
$\Delta\Gamma_s/\Gamma_s$ may well be as large as ${\cal O}(10\%)$ 
\cite{fleischer-BeLe}, thereby allowing interesting studies with 
``untagged'' $B_s$ decay rates of the kind
\begin{equation}\label{fleischer-untagged-def}
\langle\Gamma(B_s(t)\to f)\rangle\equiv \Gamma(B_s^0(t)\to f)+
\Gamma(\overline{B_s^0}(t)\to f),
\end{equation}
where we do not distinguish between initially present $B^0_s$ or
$\overline{B^0_s}$ mesons \cite{fleischer-dun-FD}. 

The focus of the following discussion will be CP violation. If we 
consider the decay of a neutral $B_q$ meson ($q\in\{d,s\}$) into a
final state $|f\rangle$, which is an eigenstate of the CP operator 
satisfying $({\cal CP})|f\rangle=\pm|f\rangle$, we obtain the following 
time-dependent CP asymmetry \cite{fleischer-RF-PHYS-REP}: 
\begin{equation}\label{fleischer-CP-asym}
\frac{\Gamma(B^0_q(t)\to f)-
\Gamma(\overline{B^0_q}(t)\to f)}{\Gamma(B^0_q(t)\to f)+
\Gamma(\overline{B^0_q}(t)\to f)}=\left[\frac{{\cal A}_{\rm CP}^{\rm dir}
\cos(\Delta M_q t)+{\cal A}_{\rm CP}^{\rm mix}
\sin(\Delta M_q t)}{\cosh(\Delta\Gamma_qt/2)-
{\cal A}_{\rm \Delta\Gamma}\sinh(\Delta\Gamma_qt/2)}\right],
\end{equation}
where
\begin{equation}\label{fleischer-obs}
{\cal A}^{\mbox{{\scriptsize dir}}}_{\mbox{{\scriptsize CP}}}
\equiv\frac{1-\bigl|\xi_f^{(q)}\bigr|^2}{1+
\bigl|\xi_f^{(q)}\bigr|^2} \quad\mbox{and}\quad
{\cal A}^{\mbox{{\scriptsize mix}}}_{\mbox{{\scriptsize
CP}}}\equiv\frac{2\,\mbox{Im}\,\xi^{(q)}_f}{1+
\bigl|\xi^{(q)}_f\bigr|^2},
\end{equation}
with
\begin{equation}
\xi_f^{(q)}=-e^{-i\phi_q}
\left[\frac{A(\overline{B_q^0}\to f)}{A(B_q^0\to f)}\right],
\end{equation}
describe the ``direct'' and ``mixing-induced'' CP-violating observables,
respectively. In the SM, the CP-violating weak $B^0_q$--$\overline{B^0_q}$ 
mixing phase $\phi_q$ is associated with the well-known box diagrams, and 
is given by
\begin{equation}\label{fleischer-phiq-def}
\phi_q=2\,\mbox{arg}(V_{tq}^\ast V_{tb})
=\left\{\begin{array}{cc}
+2\beta & \mbox{($q=d$)}\\
-2\lambda^2\eta & 
\mbox{($q=s$),}
\end{array}\right.
\end{equation}
where $\beta$ is the usual angle of the UT. Looking at 
(\ref{fleischer-CP-asym}), we observe that $\Delta\Gamma_q$ provides 
another observable ${\cal A}_{\rm \Delta\Gamma}$, which is, however, not 
independent from those in (\ref{fleischer-obs}).

The preferred mechanism for new physics (NP) to manifest itself in 
(\ref{fleischer-CP-asym}) is through contributions to 
$B^0_q$--$\overline{B^0_q}$ mixing, which is a CKM-suppressed, 
loop-induced, fourth-order weak-interaction process within the SM. 
Simple dimensional arguments suggest that NP in the TeV regime may 
well affect the $\Delta M_q$, as well as the $\phi_q$. If NP enters 
differently in $\Delta M_d$ and $\Delta M_s$, the determination
of $R_t$ from $\Delta M_d/\Delta M_s$ would be affected. On the other 
hand, NP contributions to $\phi_q$ would affect the mixing-induced 
CP asymmetries ${\cal A}_{\rm CP}^{\rm mix}$. Scenarios of this kind
were considered in several papers; for a selection, see 
\cite{fleischer-NiSi}--\cite{fleischer-FIM}. Thanks to the ``golden''
mode $B_d\to J/\psi K_{\rm S}$, direct measurements of $\sin\phi_d$ 
are already available. The current world average is given by
$\sin\phi_d\sim 0.734$, which is in accordance with the indirect range 
following from the ``CKM fits'' \cite{fleischer-CKMyellow2002}. Despite 
this remarkable feature, NP may still hide in the experimental value for 
$\sin\phi_d$, since it implies $\phi_d\sim 47^\circ \lor 133^\circ$, where 
the former solution would be consistent with the SM, while the second would 
require NP contributions to $B^0_d$--$\overline{B^0_d}$ mixing. In order to 
explore these two solutions further, we may complement them with CP violation 
in $B_d\to\pi^+\pi^-$ \cite{fleischer-FlMa2}. Following these lines 
\cite{fleischer-FIM}, we obtain an allowed region in the 
$\overline{\rho}$--$\overline{\eta}$ plane that is consistent with the 
SM for $\phi_d\sim 47^\circ$. In the case of $\phi_d\sim 133^\circ$, we 
arrive at a range in the second quadrant, which corresponds to $\gamma>90$
and is consistent with the $\varepsilon_K$ hyperbola. Interestingly, 
also this exciting possibility cannot be discarded. The current 
$B_d\to\pi^+\pi^-$ data do not yet allow us to draw definite 
conclusions, but the situation will significantly improve in the future. 
As far as $B_s$ decays are concerned, the burning question in this context 
is whether $\phi_s$, which is tiny in the SM, as can be seen in 
(\ref{fleischer-phiq-def}), is made sizeable through NP effects. In order 
to address this issue, the $B_s\to J/\psi\phi$ channel plays the key r\^ole.

\section{$B_s\to J/\psi\phi$}\label{fleischer:sec-BsPsiPhi}
This decay is the counterpart of $B_d\to J/\psi K_{\rm S}$, and exhibits an 
analogous amplitude structure:
\begin{equation}
A(B_s\to J/\psi\,\phi)\propto \left[1+ \lambda^2 ae^{i\vartheta}
e^{i\gamma}\right].
\end{equation}
Here $\gamma$ is the usual angle of the UT, and the hadronic parameter
$ae^{i\vartheta}$ measures the ratio of penguin to tree contributions, 
which is na\"\i vely expected to be of ${\cal O}(\overline{\lambda})$,
where $\overline{\lambda}={\cal O}(\lambda)={\cal O}(0.2)$ is a 
``generic'' expansion parameter \cite{fleischer-FM-NP}. In contrast
to $B_d\to J/\psi K_{\rm S}$, the final state of $B_s\to J/\psi\phi$
is an admixture of different CP eigenstates, which can be disentangled
through an angular analysis of the $J/\psi [\to\ell^+\ell^-]\phi [\to\ K^+K^-]$
decay products \cite{fleischer-DDF-ang}. Their angular distribution exhibits 
tiny direct CP violation, whereas mixing-induced CP-violating effects allow 
the extraction of
\begin{equation}\label{fleischer-sinphis}
\sin\phi_s+{\cal O}(\overline{\lambda}^3)=\sin\phi_s+{\cal O}(10^{-3}).
\end{equation}
Since we have $\phi_s=-2\lambda^2\eta={\cal O}(10^{-2})$ in the SM, the 
determination of this phase from (\ref{fleischer-sinphis}) is affected by
generic hadronic uncertainties of ${\cal O}(10\%)$, which may become
important for the LHC era. These uncertainties can be controlled with
the help of $B_d\to J/\psi \rho^0$ \cite{fleischer-RF-ang}.

Another interesting aspect of the $B_s\to J/\psi\phi$ angular distribution
is that it allows also the determination of $\cos\delta_f\cos\phi_s$, 
where $\delta_f$ is a CP-conserving strong phase. If we fix the 
{\it sign} of $\cos\delta_f$ through factorization, we may fix the 
sign of $\cos\phi_s$, which allows an {\it unambiguous} determination of
$\phi_s$ \cite{fleischer-DFN}. In this context, $B_s\to D_\pm\eta^{(')}$, 
$ D_\pm\phi$, ...\ decays are also interesting \cite{fleischer-RF-gam-eff-03}. 

The big hope is that experiments will find a {\it sizeable} value of 
$\sin\phi_s$, which would immediately signal NP. There are 
recent NP analyses where such a feature actually emerges, for example, 
within SUSY \cite{fleischer-CFMS}, or specific 
left--right-symmetric models \cite{fleischer-LRS}.

\section{$B_s\to K^+K^-$}\label{fleischer:sec-BsKK}
The decay $B_s\to K^+K^-$ is dominated by QCD penguins and complements 
$B_d\to\pi^+\pi^-$ nicely, thereby allowing a determination of $\gamma$ 
with the help of $U$-spin flavour-symmetry arguments 
\cite{fleischer-RF-BsKK}. Within the SM, we may write the corresponding 
decay amplitudes as follows:
\begin{equation}\label{fleischer-Bpipi-BsKK-ampl}
A(B_d^0\to\pi^+\pi^-)\propto
\left[e^{i\gamma}-d e^{i\theta}\right], \quad
A(B_s^0\to K^+K^-)\propto\left[e^{i\gamma}+
\left(\frac{1-\lambda^2}{\lambda^2}\right)d' e^{i\theta'}
\right],
\end{equation}
where the hadronic parameters $d e^{i\theta}$ and $d' e^{i\theta'}$ measure 
the ratios of penguin to tree contributions to $B_d^0\to\pi^+\pi^-$
and $B_s^0\to K^+K^-$, respectively. Consequently, we obtain 
\begin{equation}\label{fleischer-Bpipi-obs}
{\cal A}_{\rm CP}^{\rm dir}(B_d\to\pi^+\pi^-)=
\mbox{function}(d,\theta,\gamma), \quad
{\cal A}_{\rm CP}^{\rm mix}(B_d\to\pi^+\pi^-)=
\mbox{function}(d,\theta,\gamma,\phi_d)
\end{equation}
\begin{equation}\label{fleischer-BsKK-obs}
{\cal A}_{\rm CP}^{\rm dir}(B_s\to K^+K^-)=\mbox{function}(d',\theta',
\gamma), \quad
{\cal A}_{\rm CP}^{\rm mix}(B_s\to K^+K^-)=
\mbox{function}(d',\theta',\gamma,\phi_s).
\end{equation}
As we saw above, $\phi_d$ and $\phi_s$ can 
``straightforwardly'' be fixed, also if NP should contribute to 
$B^0_q$--$\overline{B^0_q}$ mixing. Consequently, 
${\cal A}_{\rm CP}^{\rm dir}(B_d\to\pi^+\pi^-)$ and 
${\cal A}_{\rm CP}^{\rm mix}(B_d\to\pi^+\pi^-)$ allow us to eliminate
$\theta$, thereby yielding $d$ as a function of $\gamma$ in a 
{\it theoretically clean} way. Analogously, we may fix  $d'$ as a 
function of $\gamma$ with the help of 
${\cal A}_{\rm CP}^{\rm dir}(B_s\to K^+K^-)$ and 
${\cal A}_{\rm CP}^{\rm mix}(B_s\to K^+K^-)$. 

If we look at the corresponding Feynman diagrams, we observe that 
$B_d\to\pi^+\pi^-$ and $B_s\to K^+K^-$ are related to each other through 
an interchange of all down and strange quarks. Because of this feature, 
the $U$-spin flavour symmetry of strong interactions implies
\begin{equation}\label{fleischer-U-rel}
d=d', \quad \theta=\theta'.
\end{equation}
Applying the former relation, we may extract $\gamma$ and $d$ from the clean
$\gamma$--$d$ and $\gamma$--$d'$ contours. Moreover, we may also determine 
$\theta$ and $\theta'$, allowing an interesting check of the second relation.
 
This strategy is very promising from an experimental point of 
view: at CDF-II and LHCb, experimental accuracies for $\gamma$ of 
${\cal O}(10^\circ)$ and ${\cal O}(1^\circ)$, respectively, are expected 
\cite{fleischer-TEV-BOOK,fleischer-LHC-BOOK,golutvin-fpcp03}. 
As far as $U$-spin-breaking corrections are concerned, they enter 
the determination of $\gamma$ through a relative shift of
the $\gamma$--$d$ and $\gamma$--$d'$ contours; their impact on the 
extracted value of $\gamma$ depends on the form of these curves, which 
is fixed through the measured observables. In the examples discussed in 
\cite{fleischer-RF-PHYS-REP,fleischer-RF-BsKK}, the result for
$\gamma$ would be very robust under such corrections.

As we have already noted, $B_s\to K^+K^-$ is not accessible at the BaBar
and Belle detectors. However, since we obtain $B_s\to K^+K^-$ from 
$B_d\to\pi^\mp K^\pm$ through a replacement of the down spectator quark 
through a strange quark, we have 
$\mbox{BR}(B_s\to K^+K^-)\approx\mbox{BR}(B_d\to\pi^\mp K^\pm)$.
In order to play with the $B$-factory data, we may then consider
\begin{equation}
H=\left(\frac{1-\lambda^2}{\lambda^2}\right)\left(\frac{f_K}{f_\pi}\right)^2
\left[\frac{\mbox{BR}(B_d\to\pi^+\pi^-)}{\mbox{BR}(B_d\to\pi^\mp K^\pm)}
\right]\sim7.5.
\end{equation}
If we use (\ref{fleischer-Bpipi-BsKK-ampl}) and (\ref{fleischer-U-rel}), 
we may write
\begin{equation}
H=\mbox{function}(d,\theta,\gamma),
\end{equation}
which complements (\ref{fleischer-Bpipi-obs}) and provides sufficient 
information to extract $\gamma$, $d$ and $\theta$ 
\cite{fleischer-RF-BsKK,fleischer-RF-Bpipi}. This approach 
was applied in the UT  analysis sketched at the end of 
Section~\ref{fleischer:sec-intro}, following \cite{fleischer-FIM}. 
Interestingly, $H$ implies also a very narrow SM ``target range'' in the 
${\cal A}_{\rm CP}^{\rm mix}(B_s\to K^+K^-)$--${\cal A}_{\rm CP}^{\rm dir}
(B_s\to K^+K^-)$ plane \cite{fleischer-FlMa2}. 
The measurement of BR$(B_s\to K^+K^-)$, which is expected to be soon 
available from CDF-II \cite{martin-fpcp03}, will already be an 
important achievement, allowing a better determination of $H$. Once also 
the CP asymmetries of this channel have been measured, we may fully exploit 
the physics potential of the $B_s\to K^+K^-$, $B_d\to\pi^+\pi^-$ system 
\cite{fleischer-RF-BsKK}.

\section{$B_s\to D_s^{(\ast)\pm}K^\mp$}\label{fleischer:sec-BsDsK}
Let us finally turn to colour-allowed ``tree'' decays of the kind 
$B_s\to D_s^{(\ast)\pm}K^\mp$, which complement $B_d\to D^{(\ast)\pm}\pi^\mp$ 
transitions: these modes can be treated on the same theoretical basis, and 
provide new strategies to determine $\gamma$ \cite{fleischer-RF-gam-ca}. 
Following this paper, we may write these modes generically as 
$B_q\to D_q \overline{u}_q$. Their characteristic feature is that both a 
$B^0_q$ and a $\overline{B^0_q}$ may decay into $D_q \overline{u}_q$, 
thereby leading to interference between $B^0_q$--$\overline{B^0_q}$ mixing 
and decay processes, involving the weak phase $\phi_q+\gamma$. In the case 
of $q=s$, i.e.\ $D_s\in\{D_s^+, D_s^{\ast+}, ...\}$ and 
$u_s\in\{K^+, K^{\ast+}, ...\}$, these interference effects are governed 
by a hadronic parameter $x_s e^{i\delta_s}\propto R_b\approx0.4$, where
$R_b\propto |V_{ub}/V_{cb}|$ is the usual UT side, and hence are large. 
On the other hand, for $q=d$, i.e.\ $D_d\in\{D^+, D^{\ast+}, ...\}$ 
and $u_d\in\{\pi^+, \rho^+, ...\}$, the interference effects are described 
by $x_d e^{i\delta_d}\propto -\lambda^2R_b\approx-0.02$, and hence are tiny. 
In the following, we shall only consider $B_q\to D_q \overline{u}_q$ modes, 
where at least one of the $D_q$, $\overline{u}_q$ states is a pseudoscalar 
meson; otherwise a complicated angular analysis has to be performed.
 
The time-dependent rate asymmetries of these decays take the same form 
as (\ref{fleischer-CP-asym}). It is well known that they allow a 
determination of $\phi_q+\gamma$, where the ``conventional'' approach 
works as follows \cite{fleischer-BsDsK,fleischer-BdDpi}: 
if we measure the observables 
$C(B_q\to D_q\overline{u}_q)\equiv C_q$ 
and $C(B_q\to \overline{D}_q u_q)\equiv \overline{C}_q$ provided by the
$\cos(\Delta M_qt)$ pieces, we may determine the following quantities:
\begin{equation}\label{fleischer-Cpm-def}
\langle C_q\rangle_+\equiv
(\overline{C}_q+ C_q)/2=0, \quad
\langle C_q\rangle_-\equiv
(\overline{C}_q-C_q)/2=(1-x_q^2)/(1+x_q^2),
\end{equation}
where $\langle C_q\rangle_-$ allows us to extract $x_q$. However, to this
end we have to resolve terms entering at the $x_q^2$ level. In the case 
of $q=s$, we have $x_s={\cal O}(R_b)$, implying $x_s^2={\cal O}(0.16)$, so 
that this may actually be possible, though challenging. On the other hand, 
$x_d={\cal O}(-\lambda^2R_b)$ is doubly Cabibbo-suppressed. Although it 
should be possible to resolve terms of ${\cal O}(x_d)$, this will be 
impossible for the vanishingly small $x_d^2={\cal O}(0.0004)$ 
terms, so that other approaches to fix $x_d$ are required
\cite{fleischer-BdDpi}. In order to extract $\phi_q+\gamma$, the 
mixing-induced observables $S(B_q\to D_q\overline{u}_q)\equiv S_q$ and 
$S(B_q\to \overline{D}_q u_q)\equiv \overline{S}_q$ associated with the
$\sin(\Delta M_qt)$ terms of the time-dependent rate asymmetry must be 
measured. In analogy to (\ref{fleischer-Cpm-def}), it is convenient to
introduce observable combinations $\langle S_q\rangle_\pm$. If we assume 
that $x_q$ is known, we may consider the quantities
\begin{eqnarray}
s_+&\equiv& (-1)^L
\left[\frac{1+x_q^2}{2 x_q}\right]\langle S_q\rangle_+
=+\cos\delta_q\sin(\phi_q+\gamma)\\
s_-&\equiv&(-1)^L
\left[\frac{1+x_q^2}{2x_q}\right]\langle S_q\rangle_-
=-\sin\delta_q\cos(\phi_q+\gamma),
\end{eqnarray}
which yield
\begin{equation}\label{fleischer-conv-extr}
\sin^2(\phi_q+\gamma)=\frac{1}{2}\left[(1+s_+^2-s_-^2) \pm
\sqrt{(1+s_+^2-s_-^2)^2-4s_+^2}\right].
\end{equation}
This expression implies an eightfold solution for $\phi_q+\gamma$. If we 
assume that $\mbox{sgn}(\cos\delta_q)>0$, as suggested by factorization, a 
fourfold discrete ambiguity emerges. Note that this assumption allows us also 
to fix the sign of $\sin(\phi_q+\gamma)$ through $\langle S_q\rangle_+$. 
To this end, the factor $(-1)^L$, where $L$ is the $D_q\overline{u}_q$ 
angular momentum, has to be properly taken into account 
\cite{fleischer-RF-gam-ca}. This is a crucial issue for the extraction of 
the sign of $\sin(\phi_d+\gamma)$ from $B_d\to D^{\ast\pm}\pi^\mp$ decays.

Let us now discuss new strategies to explore CP violation through 
$B_q\to D_q \overline{u}_q$ modes, following \cite{fleischer-RF-gam-ca}. 
If the width difference $\Delta\Gamma_s$ is sizeable, the ``untagged'' 
rates (see (\ref{fleischer-untagged-def}))
\begin{equation}\label{fleischer-untagged}
\langle\Gamma(B_q(t)\to D_q\overline{u}_q)\rangle=
\langle\Gamma(B_q\to D_q\overline{u}_q)\rangle 
\left[\cosh(\Delta\Gamma_qt/2)-{\cal A}_{\rm \Delta\Gamma}
(B_q\to D_q\overline{u}_q)\sinh(\Delta\Gamma_qt/2)\right]e^{-\Gamma_qt}
\end{equation}
and their CP conjugates provide
${\cal A}_{\rm \Delta\Gamma}(B_s\to D_s\overline{u}_s)
\equiv {\cal A}_{\rm \Delta\Gamma_s}$ and 
${\cal A}_{\rm \Delta\Gamma}(B_s\to \overline{D}_s u_s)\equiv 
\overline{{\cal A}}_{\rm \Delta\Gamma_s}$. Introducing, in analogy 
to (\ref{fleischer-Cpm-def}), observable combinations 
$\langle{\cal A}_{\rm \Delta\Gamma_s}\rangle_\pm$, we may derive the relations
\begin{equation}\label{fleischer-untagged-extr}
\tan(\phi_s+\gamma)=
-\left[\frac{\langle S_s\rangle_+}{\langle{\cal A}_{\rm \Delta\Gamma_s}
\rangle_+}\right]
=+\left[\frac{\langle{\cal A}_{\rm \Delta\Gamma_s}
\rangle_-}{\langle S_s\rangle_-}\right],
\end{equation}
which allow an {\it unambiguous} extraction of $\phi_s+\gamma$ 
if we assume, in addition, that $\mbox{sgn}(\cos\delta_q)>0$. 
Another important advantage 
of (\ref{fleischer-untagged-extr}) is that we do {\it not} have to rely on 
${\cal O}(x_s^2)$ terms, as $\langle S_s\rangle_\pm$ and 
$\langle {\cal A}_{\rm \Delta\Gamma_s}\rangle_\pm$ are proportional to $x_s$.
On the other hand, we need a sizeable value of $\Delta\Gamma_s$. 
Measurements of untagged rates are also very useful in the case of 
vanishingly small $\Delta\Gamma_q$, since the ``unevolved'' untagged rates 
in (\ref{fleischer-untagged}) offer various interesting strategies to 
determine $x_q$ from the ratio of 
$\langle\Gamma(B_q\to D_q\overline{u}_q)\rangle+
\langle\Gamma(B_q\to \overline{D}_q u_q)\rangle$ and CP-averaged rates of
appropriate $B^\pm$ or flavour-specific $B_q$ decays.

If we keep the hadronic parameter $x_q$ and the associated strong phase
$\delta_q$ as ``unknown'', free parameters in the expressions for the
$\langle S_q\rangle_\pm$, we may obtain bounds on $\phi_q+\gamma$ from
\begin{equation}
|\sin(\phi_q+\gamma)|\geq|\langle S_q\rangle_+|, \quad
|\cos(\phi_q+\gamma)|\geq|\langle S_q\rangle_-|.
\end{equation}
If $x_q$ is known, stronger constraints are implied by 
\begin{equation}\label{fleischer-bounds}
|\sin(\phi_q+\gamma)|\geq|s_+|, \quad
|\cos(\phi_q+\gamma)|\geq|s_-|.
\end{equation}
Once $s_+$ and $s_-$ are known, we may of course determine
$\phi_q+\gamma$ through the ``conventional'' approach, using 
(\ref{fleischer-conv-extr}). However, the bounds following from 
(\ref{fleischer-bounds}) provide essentially the same information 
and are much simpler to 
implement. Moreover, as discussed in detail in \cite{fleischer-RF-gam-ca}
for several examples within the SM, the bounds following from $B_s$ and 
$B_d$ modes may be highly complementary, thereby providing particularly 
narrow, theoretically clean ranges for $\gamma$. 

Let us now further exploit the complementarity between the processes
$B_s^0\to D_s^{(\ast)+}K^-$ and $B_d^0\to D^{(\ast)+}\pi^-$.
If we look at the corresponding decay topologies, we observe that
these channels are related to each other through an interchange of 
all down and strange quarks. Consequently, the $U$-spin symmetry 
implies $a_s=a_d$ and $\delta_s=\delta_d$, where $a_s=x_s/R_b$ and 
$a_d=-x_d/(\lambda^2 R_b)$ are the ratios of hadronic matrix elements 
entering $x_s$ and $x_d$, respectively. There are various possibilities 
to implement these relations \cite{fleischer-RF-gam-ca}. A particularly simple
picture emerges if we assume that $a_s=a_d$ {\it and} $\delta_s=\delta_d$, 
which yields
\begin{equation}
\tan\gamma=-\left[\frac{\sin\phi_d-S
\sin\phi_s}{\cos\phi_d-S\cos\phi_s}
\right]\stackrel{\phi_s=0^\circ}{=}
-\left[\frac{\sin\phi_d}{\cos\phi_d-S}\right].
\end{equation}
Here we have introduced
\begin{equation}
S=-R\left[\frac{\langle S_d\rangle_+}{\langle S_s\rangle_+}\right]
\end{equation}
with
\begin{equation}
R= \left(\frac{1-\lambda^2}{\lambda^2}\right)
\left[\frac{1}{1+x_s^2}\right],
\end{equation}
where $R$ can be fixed with the help of untagged $B_s$ rates through
\begin{equation}
R=\left(\frac{f_K}{f_\pi}\right)^2 \left[
\frac{\Gamma(\overline{B^0_s}\to D_s^{(\ast)+}\pi^-)+
\Gamma(B^0_s\to D_s^{(\ast)-}\pi^+)}{\langle\Gamma(B_s\to D_s^{(\ast)+}K^-)
\rangle+\langle\Gamma(B_s\to D_s^{(\ast)-}K^+)\rangle}\right].
\end{equation}
Alternatively, we may {\it only} assume that $\delta_s=\delta_d$ {\it or} 
that $a_s=a_d$, as discussed in detail in \cite{fleischer-RF-gam-ca}. 
Apart from features related to multiple discrete ambiguities, the 
most important advantage with respect to the ``conventional'' approach 
is that the experimental resolution of the $x_q^2$ terms is not required. In 
particular, $x_d$ does {\it not} have to be fixed, and $x_s$ may only enter 
through a $1+x_s^2$ correction, which can straightforwardly be determined 
through 
untagged $B_s$ rate measurements. In the most refined implementation of this 
strategy, the measurement of $x_d/x_s$ would {\it only} be interesting for 
the inclusion of $U$-spin-breaking effects in $a_d/a_s$. Moreover, we may 
obtain interesting insights into hadron dynamics and $U$-spin-breaking 
effects. 

In order to explore CP violation, the colour-suppressed counterparts
of the $B_q\to D_q \overline{u}_q$ modes are also very interesting. 
In the case of the $B_d\to D K_{\rm S(L)}$, $B_s\to D \eta^{(')}, D \phi$, ...\
modes, the interference effects between $B^0_q$--$\overline{B^0_q}$ mixing
and decay processes are governed by $x_{f_s}e^{i\delta_{f_s}}\propto R_b$.
If we consider the CP eigenstates $D_\pm$, we obtain additional interference
effects at the amplitude level, which involve $\gamma$, and may introduce
the following ``untagged'' rate asymmetry \cite{fleischer-RF-gam-eff-03}:
\begin{equation}
\Gamma_{+-}^{f_s}\equiv
\frac{\langle\Gamma(B_q\to D_+ f_s)\rangle-\langle
\Gamma(B_q\to D_- f_s)\rangle}{\langle\Gamma(B_q\to D_+ f_s)\rangle
+\langle\Gamma(B_q\to D_- f_s)\rangle},
\end{equation}
which allows us to constrain $\gamma$ through 
$|\cos\gamma|\geq |\Gamma_{+-}^{f_s}|$. Moreover, if we complement
$\Gamma_{+-}^{f_s}$ with 
\begin{equation}
\langle S_{f_s}\rangle_\pm\equiv(S_+^{f_s}\pm S_-^{f_s})/2,
\end{equation}
where $S_\pm^{f_s}\equiv {\cal A}_{\rm CP}^{\rm mix}(B_q\to D_\pm f_s)$,
we may derive the following simple but {\it exact} relation:
\begin{equation}
\tan\gamma\cos\phi_q=
\left[\frac{\eta_{f_s} \langle S_{f_s}
\rangle_+}{\Gamma_{+-}^{f_s}}\right]+\left[\eta_{f_s}\langle S_{f_s}\rangle_--
\sin\phi_q\right],
\end{equation}
where $\eta_{f_s}\equiv(-1)^L\eta_{\rm CP}^{f_s}$. This expression allows 
a conceptually simple, theoretically clean and essentially unambiguous 
determination of $\gamma$ \cite{fleischer-RF-gam-eff-03}; further 
applications, employing also $D$-meson decays into CP non-eigenstates,
can be found in \cite{fleischer-BDfr}. Since the interference effects are 
governed by the tiny parameter $x_{f_d}e^{i\delta_{f_d}}\propto -\lambda^2R_b$
in the case of $B_s\to D_\pm K_{\rm S(L)}$, 
$B_d\to D_\pm \pi^0, D_\pm \rho^0, ...$, these modes are not as promising
for the extraction of $\gamma$. However, they provide the relation
\begin{equation}
\eta_{f_d}\langle S_{f_d}\rangle_-=\sin\phi_q + {\cal O}(x_{f_d}^2)
=\sin\phi_q + {\cal O}(4\times 10^{-4}),
\end{equation}
allowing very interesting determinations of $\phi_q$ with theoretical 
accuracies one order of magnitude higher than those of
the conventional  $B_d\to J/\psi K_{\rm S}$, $B_s\to J/\psi \phi$
approaches (see Section~\ref{fleischer:sec-BsPsiPhi}). In particular, 
$\phi_s^{\rm SM}=-2\lambda^2\eta$ could be determined with only 
${\cal O}(1\%)$ uncertainty \cite{fleischer-RF-gam-eff-03}.

\section{Conclusions and Outlook}\label{fleischer:sec-concl}
The most exciting question concerning $B_s\to J/\psi\phi$ is whether this 
mode will exhibit sizeable mixing-induced CP-violating effects, thereby 
indicating NP contributions to $B^0_s$--$\overline{B^0_s}$ mixing. As we 
have seen, the $B_s$-meson system offers interesting avenues to extract 
$\gamma$. For example, we may employ $B_s\to K^+K^-$, which is governed by 
QCD penguin processes, to complement $B_d\to\pi^+\pi^-$, or may complement 
pure ``tree'' decays of the kind $B_s\to D_s^{(\ast)\pm}K^\mp$ with their 
$B_d\to D^{(\ast)\pm}\pi^{\mp}$ counterparts. Here the burning question
is whether the corresponding results for $\gamma$ will show discrepancies, 
which could indicate NP effects in the penguin sector. The exploration of
$B_s$ decays is the ``El Dorado'' for $B$-physics studies at hadron colliders.
Important first steps are already expected in the near future at run II
of the Tevatron, whereas the rich physics potential of the $B_s$-meson
system can be fully exploited by LHCb and BTeV.

%%%%%%%%%%%%%%%%%%%%%%%%%%%%%%%%%%%%%
% Label to flag the last page of your contribution
% Replace Perret by your name starting with a capital letter
%
\label{FleischerEnd}
 
\end{document}